\documentclass{article}
\usepackage{spconf,amsmath,graphicx}
\usepackage{booktabs}
\usepackage{float}
\usepackage{enumitem}
\setlist{nosep, leftmargin=14pt}

\usepackage{mwe} % to get dummy images

\title{Leveraging Multimodal Fusion for Enhanced Diagnosis of Multiple Retinal Diseases in Ultra-wide OCTA}
\name{Hao Wei$^{1,*}$\thanks{*Equal Contribution; https://github.com/hwei-hw/M3OCTA} 
\qquad Peilun Shi$^{1,*}$ 
\qquad Guitao Bai$^2$ 
\qquad \textit{Minqing Zhang}$^1$
\qquad \textit{Shuangle Li}$^{2,\dag}$  
\qquad \textit{Wu Yuan}$^{1,\dag}$\thanks{\dag Corresponding Author: wyuan@cuhk.edu.hk, 985750247@qq.com}}
\address{$^{1}$ Department of Biomedical Engineering, The Chinese University of Hong Kong, Hong Kong SAR\\
         $^{2}$ Department of Ophthalmology, Zigong First People’s Hospital, Zigong, China}

\begin{document}
%\ninept
%
\maketitle

\begin{abstract}
% Ultra-wide optical coherence tomography (UW-OCTA) is a promising imaging modality on the anterior and posterior of an eye over traditional OCTA, which enables to provision of a wide scanning range, multiple layers, narrow linewidth and high output power. However, the currently accessible datasets publicly are constrained by the scarcity of comprehensive hierarchical information and corresponding disease annotations. Consequently, we were compelled to curate the inaugural multimodal and expansive field-of-view OCTA dataset, with the aim of facilitating the utilization of multi-layer ultra-wide ocular vasculature information from OCTA within the scientific community. Based on collected dataset, we proposed the first yet novel multi-modal fusion framework for multiple ocular diseases diagnostic from 24 * 20 mm UW-OCTA images of multi-layer. Results demonstrate the effectiveness and high accuracy of multi-disease diagnoses from UW-OCTA. And dataset will be openly available at https://github.com/.

Ultra-wide optical coherence tomography angiography (UW-OCTA) is an emerging imaging technique that offers significant advantages over traditional OCTA by providing an exceptionally wide scanning range of up to 24 x 20 $mm^{2}$, covering both the anterior and posterior regions of the retina. However, the currently accessible UW-OCTA datasets suffer from limited comprehensive hierarchical information and corresponding disease annotations. To address this limitation, we have curated the pioneering M3OCTA dataset, which is the first multimodal (i.e., multilayer), multi-disease, and widest field-of-view UW-OCTA dataset. Furthermore, the effective utilization of multi-layer ultra-wide ocular vasculature information from UW-OCTA remains underdeveloped. To tackle this challenge, we propose the first cross-modal fusion framework that leverages multi-modal information for diagnosing multiple diseases. Through extensive experiments conducted on our openly available M3OCTA dataset, we demonstrate the effectiveness and superior performance of our method, both in fixed and varying modalities settings. The construction of the M3OCTA dataset, the first multimodal OCTA dataset encompassing multiple diseases, aims to advance research in the ophthalmic image analysis community. 

% Ultra-wide optical coherence tomography (UW-OCTA) is a promising imaging modality on the anterior and posterior of an eye over traditional OCTA, which enables to provision of a wide scanning range, multiple layers, narrow linewidth and high output power. However, the publicly available datasets are currently limited by the rich hierarchical information and corresponding disease annotations. Moreover, effectively utilizing the multi-layer ultra-wide ocular vasculature information from OCTA still remains underdeveloped. Further research is warranted to determine optimal techniques for integrating the multi-modal information captured through ultra-widefield UW-OCTA. Thus, we proposed the first yet novel multi-modal fusion framework for multiple ocular diseases diagnostic from 24 * 20 mm UW-OCTA images of multi-layer. Results demonstrate the effectiveness and high accuracy of multi-disease diagnoses from UW-OCTA. And dataset will be openly available at https://github.com/, constituting the first multimodal and widest field-of-view OCTA dataset that aims to advance research across the medical image analysis community. 
% Doing so will more fully harness the capabilities of this technology for enhanced ophthalmologic assessment and diagnosis. 
\end{abstract}


\section{Introduction}\label{sec:intro}
A thorough assessment of ocular vasculature is crucial for ophthalmologists to evaluate eye health. Angiography is the most effective modality for this purpose. Fundus fluorescein angiography (FFA) provides clear visualization of vessels with a wide field of view (FOV) but is limited by its invasive nature. Conventional Optical coherence tomography angiography (OCTA) overcomes the invasiveness of FFA but has a limited FOV~\cite{antcliff2000comparison,chen2022dual}. Ultra-wide OCTA (UW-OCTA) combines the advantages of a wide FOV and non-invasiveness, making it the optimal angiographic technique. It enables comprehensive and non-invasive visualization of ocular blood vessels, providing valuable information for diagnostic and therapeutic decisions related to ophthalmic diseases. 

% A wider field of view (FOV) OCTA has been proven to improve screening efficiency and reduce the probability of missed diagnoses compared with conventional OCTA images of 3 × 3 or 6 × 6 $mm^{2}$~\cite{zhang2018ultra}.  

\begin{figure}[t]
    \centering
    \includegraphics[width=0.5\textwidth]{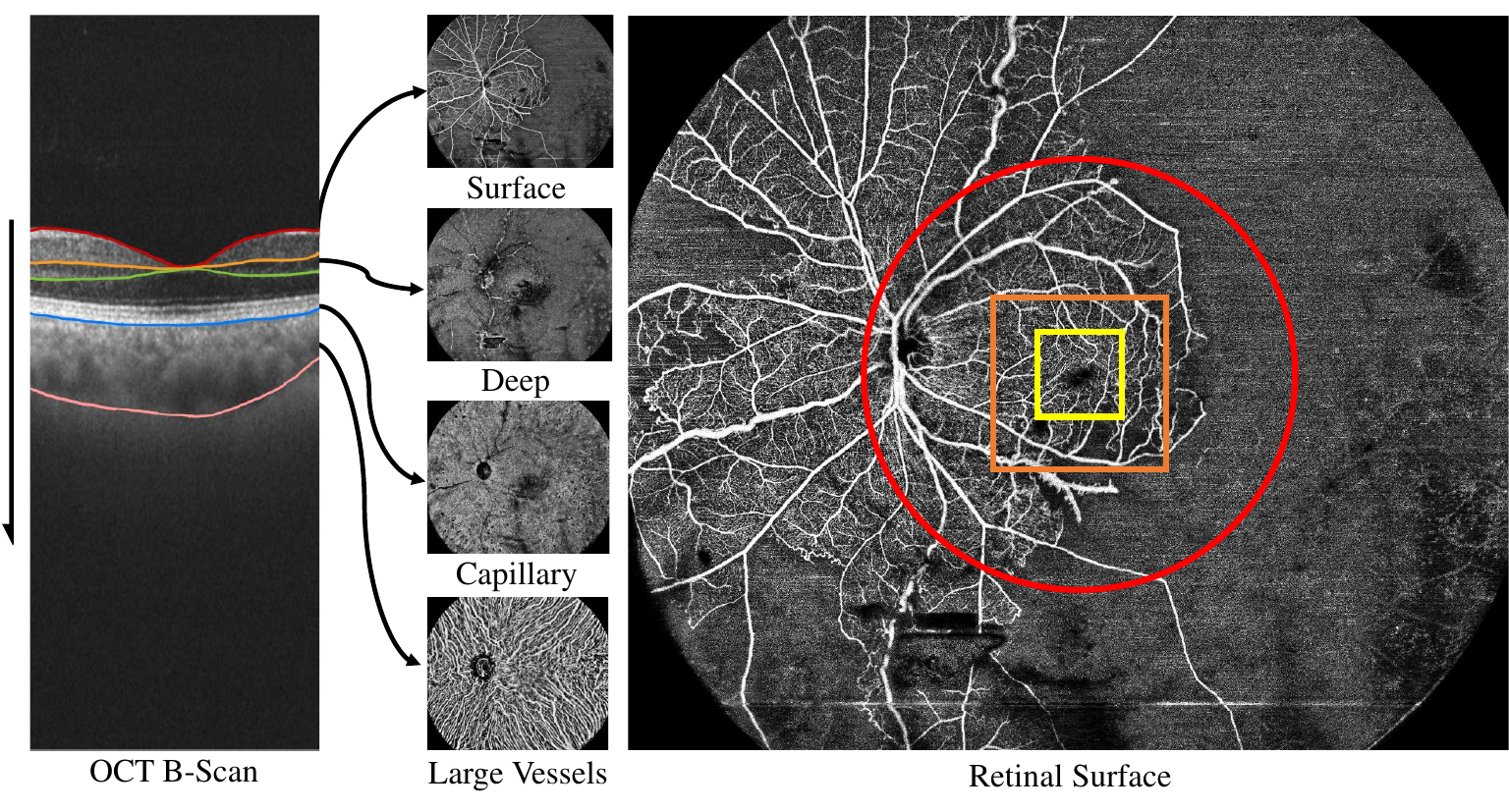}
    \caption{Illustration of proposed M3OCTA Dataset. The selected four-modal sample, scanned in 24 x 20 $mm^{2}$, contains four layer projection maps: retinal surface (inner limiting membrane-the inner plexiform layer), retinal deep (the inner plexiform layer-the outer plexiform layer), choroid capillary (bruch's membrane) and choroid large vessel (choroid layer) (The vertical arrow denotes the projection direction). The red, orange, and yellow regions denote the scan region of the regular fundus, $3\times3mm$, and $6\times6mm$ OCTA image.}
    \label{fig_region}
    \vspace{-1.0em}
\end{figure}

More recently, a new type of UW-OCTA technique (BMizar 400KHz Full-Range SS-OCT, TowardPi. Inc) affords a widest imaging range up to 24 × 20 $mm^{2}$~\cite{zheng2022advances}, as shown in Fig.\ref{fig_region}. It has been demonstrated for various clinical applications such as disease diagnosis and screening~\cite{nawaz2023unravelling,sampson2022towardsOCTA}. Projections from the three-dimensional (3D) volume in different layers show the multiple-layer visualization of both the retina and choroid, as shown in Fig.\ref{fig_region}. Considering that ocular pathologies can manifest as abnormalities across different layers of the retina, it is crucial to take into account lesions on each affected layer in order to facilitate a multi-modality-based diagnosis of diseases. Therefore, it is of great importance to develop and validate methods that effectively utilize the multi-modality information. Recently, the convolutional neural network (CNN) based two-branch paradigm~\cite{wang2022learning} or the hybrid manner of CNN with Transformer~\cite{zhang2023tformer} achieved comparable performance in multi-modal disease diagnosis. However, the parameters of these methods usually increase multiplicatively with the number of involved modalities, and the trained model cannot dynamically suit the varying modalities' inference, limiting practical usage. Moreover, the existing public OCTA datasets are limited by either the narrow scan range or the single modality.

% Consequently, developing and validating a deep-learning model to tackle the multi-modality information of UW-OCTA is significant for ophthalmologist screening and diagnostics. This underscores the crucial role that image fusion methods play in this context. Analysis from multimodal images has attracted attention in the medical domain which provides more comprehensive information for diagnostic\cite{azam2022review}. By fusing images from different modalities, clinicians can leverage the strengths of each method, overcome their individual limitations, and obtain a more holistic view of the patient's condition. Moreover, knowledge from multimodal can improve the accuracy and efficiency of disease diagnosis~\cite{lxm2020selfmultimodal}, treatment planning, and monitoring~\cite{topol2023artificial,qiu2023large}. Meanwhile, foundation models proposed recently provide a unique opportunity for advancing the analysis of ophthalmological modalities.~\cite{visionfm,zhou2023retfoundation}. Which has exhibited generalization capabilities from existing modalities, its limitations become apparent when applied to unexplored ones~\cite{shi2023generalist}.
% In this work, we proposed a multimodal fusion framework for multiple ophthalmic disease diagnoses from UW-OCTA by leveraging the foundation model with novel medical image modalities. 

To tackle these challenges, we compile the first retinal UW-OCTA dataset of widest FOV and containing five diseases and four modalities, termed as M3OCTA, and further propose a novel cross-modal fusion framework (CMF-Net) to leverage multi-modal information in this dataset for multiple diseases diagnosis in multi-label setting. Specifically, the unlabeled images and the whole train set are utilized to pre-train the ViT-based encoder by the multi-modal based masked image modeling, which learns relationships between any two modalities by the global self-attention. Then, we propose the attention-based cross-modal fusion (CMF) block to reinforce and extract the multi-modal semantics for the disease diagnosis. Moreover, our design enables a varying number of modal inputs but without performance drop during the inference stage, increasing suitability and compatibility in clinical use. In summary, our main contributions are as follows:
\begin{itemize}
    \item We introduce the first multi-modal UW-OCTA dataset with multiple disease annotations, i.e., M3OCTA, aiming to promote advances in ophthalmic image analysis.
    % \item The first multi-modal ultra-wide optical coherence tomography angiography (UW-OCTA) dataset (M3OCTA) is published, which aims to promote advances in ophthalmic and biomedical image analysis.
    \item We propose the cross-modal fusion network (CMF-Net) to involve multi-layer OCTA images for accurate and robust diagnosis of multiple retinal diseases. Its support for dynamic input modalities during inference greatly improves the suitability for the practical use.
    \item Extensive experiments have verified the effectiveness and superiority of the proposed method. Investigation of the impact on varying input modalities also proves the value of the multiple modalities of the proposed dataset. 
    % \item A multimodal fusion network architecture has been proposed to enable dynamic input for the diagnostic classification of multiple diseases. 
    % \item Evaluation of recent ophthalmology-related foundation model on UW-OCTA multi-disease diagnostic.
\end{itemize}

\section{Dataset}
% \subsection{Data}

\begin{figure}[h]
    \centering
    \includegraphics[width=0.5\textwidth]{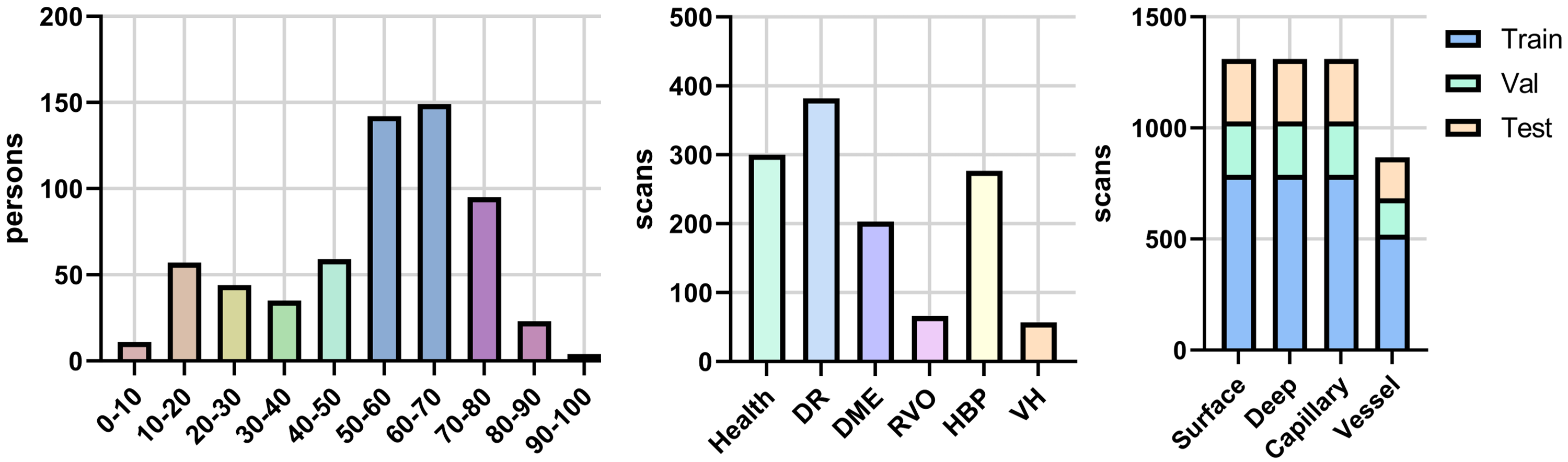}
    \caption{Age, diseases and modalities statistics of M3OCTA dataset.}
    \label{fig:enter-label}
    \vspace{-1.0em}
\end{figure}

Our proposed M3OCTA is the first multi-modal based ultra-wide retinal OCTA dataset, involving 1637 scans from 1046 eyes of 620 individuals imaged in Zigong First People’s Hospital through $24\times20$ scan mode. Specifically, 1067 scans contains choroid large vessel image; images of 1310 scans from 496 people are labeled as six classes in multi-label setting, including healthy, diabetic retinopathy (DR), diabetic macular edema (DME), Retinal Vein Occlusion (RVO), Hypertension (HBP) and Vitreous Hemorrhage (VH), and then split into train, validation and test set as 6:2:2. The remaining unlabeled data are only used in the pretraining step. Details of our M3OCTA and other public ones are listed in Table.\ref{tab_dataset}. Compared with others, M3OCTA dataset demonstrates superiorities in several aspects including the number of modalities, number of patients, image resolution, and FOV. 

% In this collection of UW-OCTA data, the focus of the analysis revolves around the ocular region, with a comprehensive assessment of 1046 eyes of 620 individuals. Totally, a thorough examination of 1637 scans has been conducted. Six types of eye stages are involved including healthy, diabetic retinopathy (DR), diabetic macular edema (DME), Retinal Vein Occlusion (RVO), Hypertension (HBP) and Vitreous Hemorrhage (VH). Images of 496 people were labeled and have been used as labels for train, validation, and testing, the rest of unlabeled data were only used during pretraining. Details of the data are listed in Table.\ref{tab_dataset}. In comparison to the published OCTA dataset, the proposed MM-OCTA dataset demonstrates superiorities in several aspects including the number of modalities, number of patients included, resolution, and field of view (FOV). However, due to the constraints in the usage time of UW-OCTA, the disease diversity covered in the MM-OCTA dataset is not as extensive as the proven OCTA dataset. 

% Including the age distribution of the patient, disease distribution as well as modality distribution.

\begin{table}[]
\centering
\caption{Summary of the public OCTA datasets and ours, where $R$ and $C$ denotes the retina and choroid}
\label{tab_dataset}
\resizebox{\linewidth}{!}{
\begin{tabular}{lcccccc}
\hline
Dataset    & Modalities & Subjects       & Diseases        &Regions         & Resolution                                               & FOV(mm)                                            \\ \hline
Giarrarano et al.~\cite{giarratano2020gocta} & 1          & 11             & 1              &R          & 91 × 91                                                     & 3 × 3                                               \\
ROSE~\cite{ma2020rose}       & 1          & 151            & -              &R          & \begin{tabular}[c]{@{}c@{}}304 × 304\\ 512 × 512\end{tabular} & 3 × 3                                                \\
OCTAGON~\cite{diaz2019OCTAGON}    & 1          & 213            & 2              &R          & 320 × 320                                                   & \begin{tabular}[c]{@{}c@{}}6 × 6 \\ 3 × 3\end{tabular} \\
FAZID~\cite{agarwal2020fazid}      & 1          & 304            & 3              &R          & 420 × 420                                                   & 6 × 6                                                \\
OCTA-500~\cite{li2020octa500}   & 2          & 500            & \textbf{\textgreater 12} &R  & 640 × 400                                                   & \begin{tabular}[c]{@{}c@{}}6 × 6 \\ 3 × 3\end{tabular} \\
DRAC~\cite{hou2022drac}       & 1          & \textless{}611 & 1             &R           & 1024 × 1024                                                 & 12 × 12                                              \\ \hline
\textbf{M3OCTA}    & \textbf{4} & \textbf{1067}  & 5             &\textbf{R \& C }          & \textbf{1536 × 1280}                                        & \textbf{24 × 20}                                     \\ \hline
\end{tabular}}
\vspace{-1.0em}
\end{table}

\section{Method}
% The wealth of information provided by UW-OCTA across multiple retinal layers warrants a diagnostic approach that comprehensively accounts for and integrates these multilayer data. To this end, 

\begin{figure}
    \centering
    \includegraphics[width=0.48\textwidth]{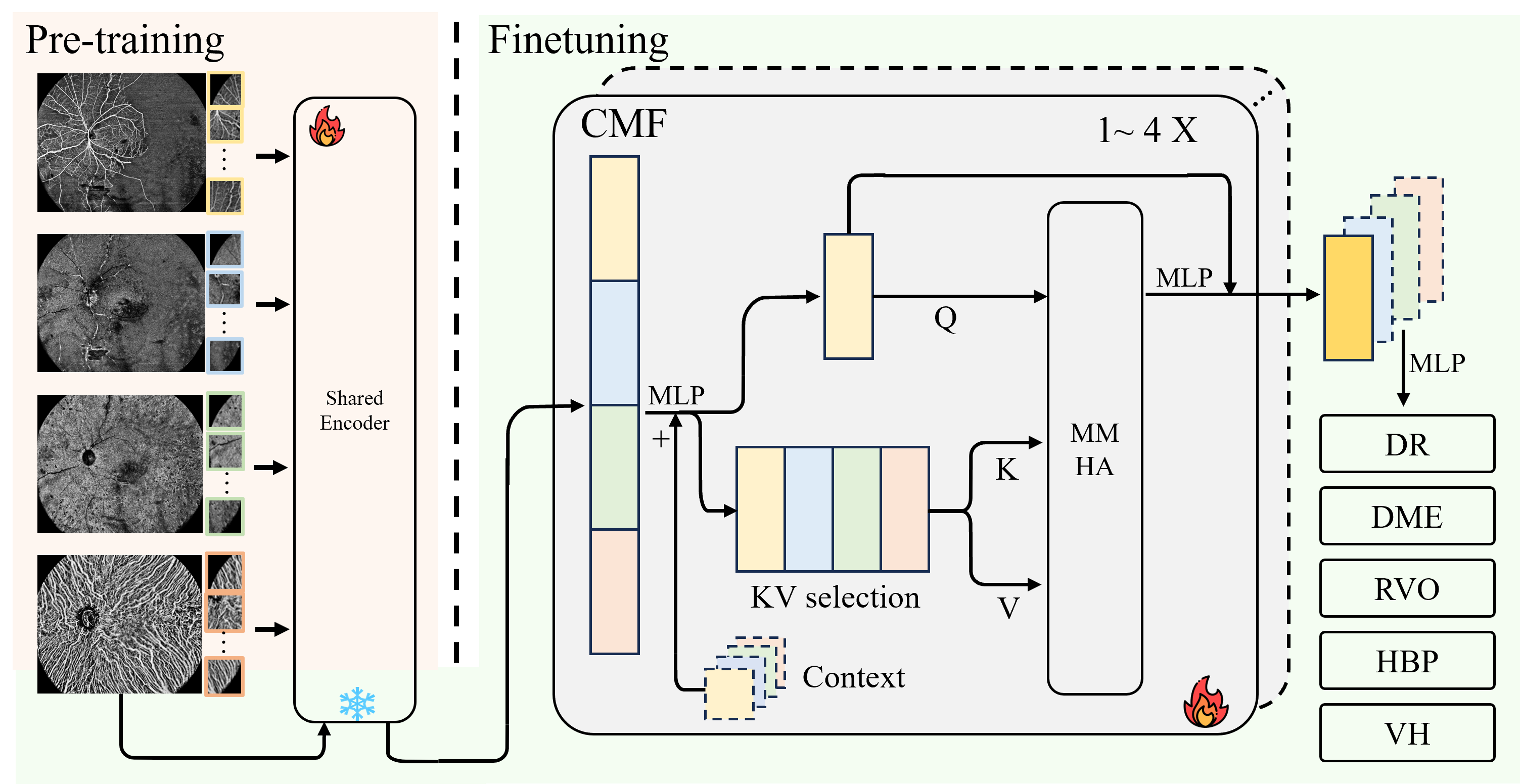}
    \caption{Our proposed framework for the cross-modal fusion and multiple retinal diseases diagnosis, including two steps: 1) encoder pretraining and 2) decoder finetuning. We first pre-train the vision transformer through the masked image modeling learning paradigm to learn multi-modal feature presentations. Then, freezing the encoder and adopting the proposed cross-modal fusion (CMF) block for each modality (share weights) to fuse the multi-modal features and extract the high-level semantics for the following diagnosis. During inference, our method can dynamically process inputs of varying modal numbers and provide stable and comparable performances. }
    \label{fig_method}
    \vspace{-1.0em}
\end{figure}

We design a novel framework that leverages the pre-trained transformer encoder and fuses multi-modal information for diagnostics, as shown in Fig. {\ref{fig_method}}. The proposed framework is divided into two stages: transformer encoder pretraining and multi-modal fusion.

% In this study, we mainly focus on the second stage, to design powerful multi-modal fusion strategies

\subsection{Multi-modal Encoder Pretraining}

% \begin{figure*}
%     \centering
%     \includegraphics[width=\textwidth]{Figs/method2.png}
%     \caption{Our proposed framework for the cross-modal fusion and multiple retinal diseases diagnosis, including two steps: 1) encoder pretraining and 2) decoder finetuning. We first pre-train the vision transformer through the masked image modeling learning paradigm to learn multi-modal feature presentations. Then, freezing the encoder and adopting the proposed cross-modal fusion (CMF) blocks to fuse the multi-modal features and extract the high-level semantics for the following diagnosis. During inference, our method can dynamically process inputs of varying modal numbers and provide stable and comparable performances. }
%     \label{fig_method}
%     \vspace{-1.0em}
% \end{figure*}

This stage follows the paradigm of masked image modeling~\cite{bachmann2022multimae}: masks the patch of input images are used to train an encoder to recover the patch for better multi-modal feature representations. Specifically, we first build the pretraining set to include all the unlabeled samples and the whole train set. Then, the multi-modal UW-OCTA images are split into image patches, and $25\%$ of these patches are randomly selected as the inputs while the remaining ones would be utilized as the targets to train the encoder~\cite{he2022masked}. During this selection process, we adopt the Dirichlet distribution to assign the number of input patches in each modality. In addition, for samples with missing modalities, we introduce dynamic controls after Dirichlet sampling to set the selected proportion of missing modalities as zeros and also increase the sampled patches in other modal images to achieve the fixed number of input patches. 

Following MAE~\cite{he2022masked}, four decoders are designed to reconstruct four modalities, where each decoder includes a linear projection layer to reduce the dimensions, positional embedding, and transformer blocks. For those input samples with missing modality, the missing one will not contribute to the loss of reconstruction. Through this reconstruction process, the encoder can model global and long-range interactions between any two modalities, which would be particularly beneficial in the multi-modal based diagnosis.   

% Following the paradigm of masked autoencoder (MAE)~\cite{he2022masked}. The proposed model architecture employs an encoder-decoder structure. A batch of multi-layer UW-OCTA images of different modalities is encoded by a frozen encoder. The full set of visible tokens from the encoder feature representations, after linear projection, are input to each decoder. To integrate information from the encoded tokens of other UW-OCTA layers, we incorporate a fusion layer in each decoder. The tokens from other modalities serve as queries, while all encoded tokens are utilized as keys and values shown in Fig.3. Prior to this fusion, sine-cosine positional embeddings and learned modality embeddings are added to the tokens. The fusion layer is followed by a small multilayer perceptron and Transformer blocks in each decoder. This architecture enables contextual integration of multimodal medical image representations during the decoding process.

\subsection{Cross-modal Fusion Decoder}

% Although multi-modal image fusion has been achieved within the pretraining step, 
% Note -- This fusion is achieved within the encoder, and then the cross-attention in the decoder reinforces the fused information for retrieval.
% After pretraining, the cross-modal fusion (CMF) block (shown in Fig.\ref{fig_method}) is designed to reinforce the multi-modal representations for better diagnosis performance. 

To reinforce the multi-modal representations and extract high-level semantics, we design the cross-modal fusion (CMF) block, as shown in Fig.\ref{fig_method}, and then parallel (but share weights) this block for each modality to build the diagnosis decoder. Specifically, a linear projection layer is incorporated to tailor the encoder outputs to the decoder's dimension. Following this projection, the decoder inputs are enhanced by the addition of both sin-cosine positional embeddings and modality-specific embeddings (context) that have been previously learned. Subsequently, the process continues with a masked multi-head attention (MMHA) layer, a Multilayer Perceptron (MLP), and the final classifier. 

Based on the mentioned process, for the $M$ modal encoded features $z_0$ from the encoder:
% \begin{equation}
\begin{align}
z_0 & = [CLS; x_0^0;x_1^0;,,,x_P^0;x_0^1;...;x_P^M], P=196 \\
z_{proj} & = proj(z_0) + E_{pos} + [E^0;...E^M] \\
Q^i & = [x_0^i;x_1^i;,,,x_P^i],  x_j^i \in  z_{proj} , i = {0,...,M} \\
KV & = \sum_{i=0}^{M}Q^i  \\
z^i & = MMHA(Q^i, KV, mask_i), i = {0,...,M} \\
z_{fuse}^i & = z^i + MLP(z^i) +Q^i, i = {0,...,M}
\end{align}
where $E_{pos}$ and $E^{i}$ denote the sin-cosine positional embeddings and modality $i$ embeddings. To learn the relationships between any modalities, We utilize the tokens of each modality as the $key$ and the summation of all that of each modality as the $key-value$ pair and then apply the multi-head attention to interact with each other, where other implementations about $key-value$ pair are discussed in the following ablation study. $mask_i \in \{0,1\}$ represents whether the modality $i$ in each input sample is missing (0) or not (1). The corresponding position of the attention computation in the MMHA layer will be set to negative infinity before the $softmax$ operation. Finally, all fused features $z_{fuse}$ are concatenated together as the inputs of the classifier.

% \end{equation}

% \begin{figure}
%     \centering
%     \includegraphics[width=0.5\textwidth]{Figs/method.png}
%     \caption{Caption}
%     \label{fig:enter-label}
% \end{figure}

\section{Experiment}
In order to investigate the capabilities of our proposed method, we performed extensive experiments utilizing the datasets with multimodal fusion algorithms. Furthermore, we evaluated foundation models trained on natural and retinal-related images, as feature extractors in an encoder role.

\subsection{Implement Details and Evaluation Metrics}
The proposed method was implemented in Pytorch using a ViT-B~\cite{dosovitskiy2020vit} with 16×16 pixel patches as the backbone encoder, and input images with the size of $224 \times 224$. During pretraining, random cropping and random horizontal flipping are used as the data augmentations, with the Cosine Anneal scheduler and AdamW~\cite{loshchilov2017decoupled} optimizer ($lr=0.001$, warm up= 40 with total = 1600 epochs, batch size = 160) and MSE loss. For the finetuning stages, the decoder was trained using an initial learning rate of $2.5e-4$, binary cross entropy loss, with the AdamW optimizer for 100 epochs and batch size = 128. The whole approach was implemented and trained using Pytorch on NVIDIA 4090 GPUs. 

% The proposed method was implemented in Pytorch using a ViT-B~\cite{dosovitskiy2020vit} with 16×16 pixel patches as the backbone encoder, and all the input images with the size of $224 \times 224$. During pretraining, random cropping and random horizontal flipping are used as the data augmentations, with the Cosine Anneal scheduler ($lr=0.001$, warm up= 40 with total = 1600 epochs) and AdamW optimizer (batch size = 160). For the finetuning stages, more data augmentation techniques were involved compared with the pretraining stage, including random cropping, random affine transformations, color jittering, horizontal flipping, and color normalization. The decoder was trained using an initial learning rate of $2.5e-4$, with the AdamW optimizer for 100 epochs and batch size = 128. The whole approach was implemented and trained using Pytorch on NVIDIA 4090 GPUs. 
% To stabilize optimization, the Cosine Anneal schedule was used to decay the learning rate after the warm-up of 5 epochs. MSE and Binary Cross entropy are used in the pretraining and finetuning stages, respectively. The whole approach was implemented and trained using Pytorch on NVIDIA 4090 GPUs.

Accuracy (ACC), Area Under the Receiver Operating Characteristic (AUROC), Average-precision (AP), and F1-score (F1) are adopted in the multi-label setting to evaluate all involved experiments by using the TorchMetrics library~\cite{torchmetrics}.
% During training, multiple data augmentation techniques were utilized to generate varied views of each input image, including random cropping, random affine transformations, color jittering, horizontal flipping, and color normalization. The fusion network was trained using an initial learning rate of 2.5e-4, with the AdamW optimizer for 100 epochs. To stabilize optimization, a Warmup learning rate schedule was applied. Thereafter, a Cosine Anneal schedule smoothly decayed the learning rate. This combined scheduling enabled larger update steps during early training, while allowing fine-tuning of the weights towards convergence. The whole approach was implemented and trained using Pytorch on NVIDIA 4090 GPUs.

% \subsection{valuation Metrics}
% Prevailing evaluation protocol has been employed in our experiments including Acc, AUC, F1 Score and Recall.

% Where TP, FN, FP, TN refer to True Positive, False Negative, False Positive and True Negative respectively. 

\subsection{Comparison}

To evaluate and compare the performance and effectiveness of the proposed framework, we select the most commonly used classification network: ResNet-18~\cite{he2016resnet}, ConvNeXT\cite{liu2022convnet} and multi-modal image classification network in the medical domain: TFormer~\cite{zhang2023tformer}, MMC\cite{wang2022learning}. All methods are trained and tested on all four modalities except the TFormer and MMC, which natively and only support two-modal images (retinal surface and deep layer images are used in our experiments). Table \ref{comparison} lists all results of compared methods, where the lower boundary of theoretical performance by random predictions is shown in the first row. Compared with other approaches, our network achieves state-of-the-art (SOTA) performance across all evaluation metrics in four-modal and two-modal settings.

% 1. We selected the most classic classification network, ResNet~\cite{he2016resnet} MobileNet~\cite{howard2019mobilev3} et al., as well as medical multimodal fusion networks~\cite{zhang2023tformer} for comparison. The results, as shown in the Fig~\ref{comparison} demonstrate that our network achieves superior performance across all evaluation metrics compared to other approaches. The lower boundary of theoretical performance has been proved in the first row of Fig~\ref{comparison} with a random setting. We also follow the initial setting of MMC and TFormer in which input modality requires 2 to compare with our method with 2 inputs. 
% \usepackage{booktabs}

\begin{table}[h]
\centering
\caption{The classification results comparison between our method and others. Where $*$ denotes method was trained and tested on two modalities (retinal surface and deep images).}
\resizebox{0.45\textwidth}{!}{
\begin{tabular}{@{}lcccc@{}}
\toprule
Method  & ACC & AUROC & AP & F1 \\ \midrule
Random  & 50.39    & 48.90    & 19.43   & 25.56\\
Resnet\cite{he2016resnet} & 81.85    & 82.17    & 49.91   & 37.65 \\
Convnext-Tiny\cite{liu2022convnet} & 83.20    & 83.68    & 54.65   & 50.88\\ \hline
MMC*\cite{wang2022learning} & 84.27    & 84.26    & 50.69   & 46.21 \\
TFormer*\cite{zhang2023tformer} & 84.41    & 69.11    & 37.21   & 50.48  \\
Ours* & 83.49    & 84.88    &  56.44  & 51.52 \\ \hline
Ours    & \textbf{84.98}    & \textbf{84.77}    & \textbf{59.99}   & \textbf{59.60} \\ \bottomrule
\end{tabular}}
\label{comparison}
\vspace{-1.0em}
\end{table}

% 2. Single modality and Multi modality Test\\

To validate the performance gain brought by the multiple modalities, we pre-train another encoder on only one modality, i.e., the retinal surface image, and then finetune the decoder on these two encoders by 1 (retinal surface), 2 (retinal surface + deep), 3 (retinal surface + deep + choroid Capillary), or all four modalities. From the results in Table.\ref{finetuned_score}, the 4-modal pretrained encoder outperforms the 1-modal encoder in all 4 scenarios with an average improvement of $10.9$, which demonstrates the effectiveness of additional modalities. For the 4-modal encoder-based experiments, the performance gain is observed as the number of input modalities increases. More comprehensive details brought by more modalities boost the performance of our model. However, the opposite trend is found in another group, which may result from the weak multi-modal features of the encoder. In summary, we determined that our method attains an optimal solution as the number of input modalities increases and multi-modal images can boost diagnosis performance. 

% Two pre-trained models based on both single and four modalities (refer to column) with test performances have been listed in Table~\ref{F1 score}. As our model is trained on multimodal data and can accommodate flexible modal inputs, the number of test modalities can vary. Consequently, we examined the influence of the number of input modalities on prediction accuracy. The table below showcases the results of the tests, encompassing single to multiple modal inputs. As the number of input modalities was increased, the performance of the model was likewise enhanced. It should be noted the numerical size represents the cumulative degree of the UW-OCTA image from the surface to the deep layer, rather than randomly fusing the images of the specified numbers. Empirical results demonstrate the superiority of our proposed method over the comparative approach when utilizing consistent modal data. Additionally, we determined that our method attains an optimal solution as the number of input modalities increases. 

\begin{table}[h]
\centering
\caption{Influence of modality variation on performance (F1 Score) during pretraining and finetuning stages}
% \caption{The impact of varying involved modalities on performance (F1) in the pretraining and finetuning stages.}
\resizebox{\linewidth}{!}{
\begin{tabular}{@{}ccccc@{}}
\toprule
Modality Number & 1 & 2 & 3 & 4 \\ \midrule
Pretrained on 1 Modality  &  46.48   & 44.85    & 44.57   &  41.72      \\
Pretrained on 4 Modality   & \textbf{52.73}    & \textbf{51.52}    & \textbf{57.37}   &  \textbf{59.60}      \\ \bottomrule
\end{tabular}}
\label{finetuned_score}
\vspace{-1.0em}
\end{table}

\begin{figure}
    \centering
    \includegraphics[width=0.5\textwidth]{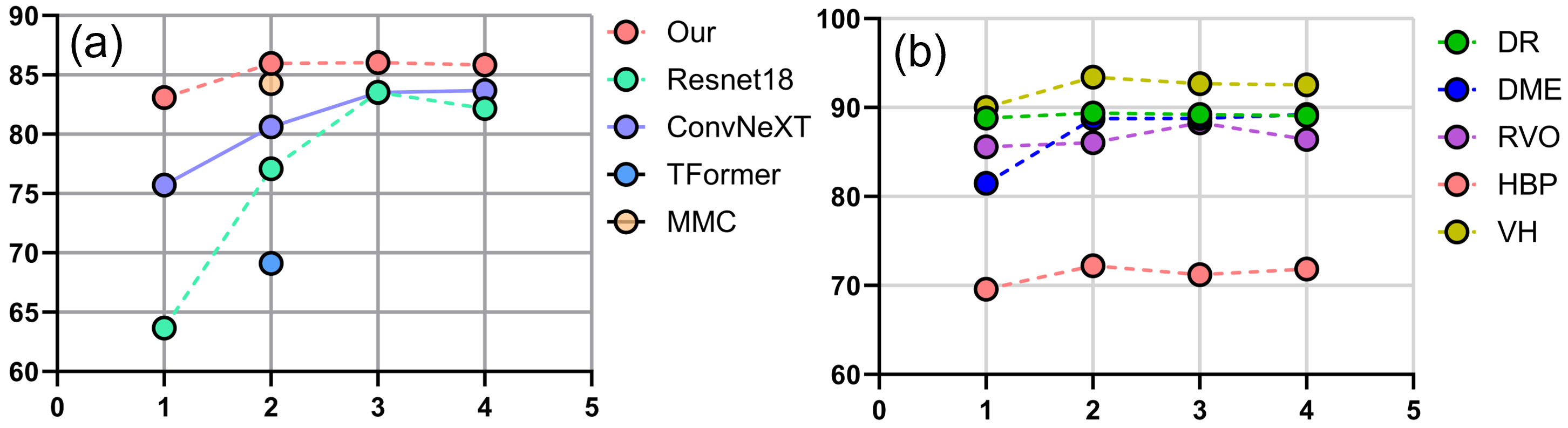}
    \caption{Stability of model performance (F1-Score) versus variation number of test modalities (arranged in order from retinal surface to choroid vessels) for different models (a) and five diseases (b)}
    % \caption{The performance (F1-score) stability of our model when test modalities vary, where the x-axis denotes the number of involved modalities according to the order from retinal surface to choroid vessels.}
    \label{Test_AUC_F1}
    \vspace{-1.0em}
\end{figure}

We subjected our approach (which incorporates a 4-modal fine-tuned decoder with a 4-modal pre-trained encoder) to an evaluation across diverse modalities for validating stability when facing varying modalities shown in Fig.\ref{Test_AUC_F1}(a). With the number of input modalities increases, the performance gain is also observed in almost all comparable methods, where our model shows more stability than others, demonstrating the effectiveness of learned multi-modal features in our design. Further, this stability also can be observed in all five disease types in Fig.\ref{Test_AUC_F1}(b).

% Considering the support of dynamic input modalities in our method, performance stability when facing varying modalities is also important. To validate it, We test our method (4-modal finetuned decoder with the 4-modal pretrained encoder) on varying modalities, as shown in Fig.\ref{Test_AUC_F1}(a). As the number of input modalities increases, the performance gain is also observed in almost all comparable methods, where our model shows more stability than others, demonstrating the effectiveness of learned multi-modal features in our design. Further, this stability also can be observed in all the five classes in Fig.\ref{Test_AUC_F1}(b).

\subsection{Ablation Study}
We also conduct ablation experiments to demonstrate the influence of pretraining datasets. Specifically, we alternately replaced the encoders from pre-trained or foundation models which are listed in Table~\ref{Weight}. The experimental results demonstrate that the greater the similarity between pretrained and target images (smaller domain gap\cite{wei2023caudr}), the better the performance~\cite{shi2023generalist}. In addition, different implementations in Eq.(4) also have an impact on the final results. In Table.\ref{Weight}, V3 achieve the best performance.

% We also conduct ablation experiments to demonstrate the influence of pretraining datasets. Specifically, we alternately replaced the encoders from pre-trained or foundation models which are listed in Table~\ref{Weight}, which are pretrained on natural images, retinal fundus~\cite{zhou2023retfoundation}, retinal fluorescein fundus angiography\cite{visionfm}, and OCTA (ours). The experimental results demonstrate that the greater the similarity between pretrained and target images, the better the performance. In addition, different implementations in Eq.(4) also have an impact on the final results. In Table.\ref{Weight}, V3 (addition) achieve the best performance compared with V1 (randomly sampling and addition) and V2 (concatenation).

\begin{table}[H]
\vspace{-1.0em}
\caption{Ablation studies on pretrained images and implementations of KV in Eq.(4). V1 refers to randomly sampling and addition, V2 refers tp concatenation, and V3 refers to addition.}
\label{tab:my-table}
\resizebox{\linewidth}{!}{
\begin{tabular}{lcccc}
\hline
Weights         & ImageNet\cite{bachmann2022multimae} & RETFound~\cite{zhou2023retfoundation} & VisionFM~\cite{visionfm}       & Ours           \\ \hline
Images    & Natural & Fundus & FFA & UW-OCTA \\
F1-score             & 21.66    & 31.43    & 37.89          & \textbf{59.60} \\ \hline
Implementation & V1     & V2       & V3             & -              \\ \hline
F1-score      & 55.42    & 57.98    & \textbf{59.60} & -              \\ \hline
\end{tabular}}
\label{Weight}
\vspace{-1.0em}
\end{table}

%3. To further validate the effectiveness of our proposed fusion method, we alternately replaced the encoders from various pre-trained or foundation models which are listed in Table~\ref{Weight}. The experimental results demonstrate that our fusion method achieves optimal performance, thereby indicating that the proposed fusion method is adaptable to multiple pre-trained models.

% \subsection{Ablation Study}
% To investigate the interrelationships among multimodal information within the context of multi-disease diagnosis, as well as each individual modality and each disease.  We have enriched our study with additional experiments. These include twenty distinct sets of results that correlate four modalities with five diagnoses, accompanied by an analysis of the mean value. For multi-disease diagnostic under an overlap scenario, it is certain that more modality information achieves higher performance than a single modality input.

% 1. Ablation of different implementations of key-value

% 2. The performance of different classes vs different modalities num.

% \input{discussion}
\section{Conclusion}

In this study, we first introduce a new multi-modal ultra-wide retinal OCTA dataset (M3OCTA) with the multi-diseases annotations. Then the cross-modal fusion network (CMF-Net) is proposed to leverage multi-modal OCTA images for efficiently diagnosing a range of ophthalmological diseases. Moreover, CMF-Net enables a varying number of input modalities without performance drop during the inference phase, which greatly increases the applicability and compatibility in different clinical scenarios. Both quantified and visualized results indicate that our model exhibits robust performance. Our fusion method successfully provides a new view for leveraging the multi-modalities to innovative ophthalmic imaging technology. Beyond the fusion method, we delved deeper into the potential application of retina-related foundational models on UW-OCTA. Our findings also suggest that incorporating more information derived from multi-modalities for foundational models is also essential for robust performance. 

% Our findings suggest that to fully harness the potential of artificial intelligence within the field of ophthalmology, it is essential to incorporate novel information derived from innovative modalities with established foundational models.

% In this study, we introduce a dynamic multi-modal fusion method capable of handling various layers from UW-OCTA. By utilizing the encoder from the pre-trained foundational model, the comprehensive framework is capable of efficiently diagnosing a range of ophthalmological diseases. The quantified and visualized results both indicate that the model exhibits robust performance. Our fusion method successfully provides a new view for leveraging the foundation model to innovative medical imaging technology.
\section{ACKNOWLEDGEMENT}
This work was supported in part by the Research Grants Council (RGC) of Hong Kong SAR (ECS24211020, GRF1420\\3821, GRF14216222), the Innovation and Technology Fund (ITF) of Hong Kong SAR (ITS/240/21), the Science, Technology and Innovation Commission (STIC) of ShenzhenMunicipality (SGDX20220530111005039)
\section{ COMPLIANCE WITH ETHICAL STANDARDS}
% This work was supported in part by the Research Grants Council (RGC) of Hong Kong SAR (ECS24211020, GRF142-03821, GRF14216222), the Innovation and Technology Fund (ITF) of Hong Kong SAR (ITS/240/21), the Science, Technology and Innovation Commission (STIC) of Shenzhen Municipality (SGDX20220530111005039), *Corresponding authors (WY): wyuan@cuhk.edu.hk

This study was performed in line with the principles of the Declaration of Helsinki and approved by the local institutional review board. Informed written consent was obtained from all institutional patients.

% \subsection{Info}
% 1.MATR: Multimodal Medical Image Fusion via Multiscale Adaptive Transformer TIP 2022\\https://github.com/tthinking/MATR\\
% 2.LRRNet: A novel representation learning guided fusion framework for infrared and visible images TPAMI 2023\\https://github.com/hli1221/imagefusion-LRRNet\\
% 3.Self-supervised multi-modal fusion network for multi-modal thyroid ultrasound image diagnosis\\
% 4.Skin lesion classification based on two-modal images using a multi-scale fully-shared fusion network\\https://github.com/pixixiaonaogou/MLSDR\\
% 5.TFormer: A throughout fusion transformer for multi-modal skin lesion diagnosis\\https://github.com/zylbuaa/TFormer\\

% \subsection{baseline}
% 1.Multi-modal Pathological Pre-training via Masked Autoencoders for Breast Cancer Diagnosis

\bibliographystyle{IEEEtran}

\bibliography{reference}

\end{document}